\documentstyle[psfig,draft]{mn}

\def\gs{\mathrel{\raise0.35ex\hbox{$\scriptstyle >$}\kern-0.6em
\lower0.40ex\hbox{{$\scriptstyle \sim$}}}}
\def\ls{\mathrel{\raise0.35ex\hbox{$\scriptstyle <$}\kern-0.6em
\lower0.40ex\hbox{{$\scriptstyle \sim$}}}}

\title[A submillimetre-selected hyperluminous galaxy]
      {A hyperluminous galaxy at {\bf{\em z}} = 2.8 found in a deep
       submillimetre survey}

\author[R.\,J.\ Ivison et al.]
       {R.\,J.\ Ivison,$^{\! 1}$ Ian Smail,$^{\! 2}$ J.-F.\ Le Borgne,$^{\!
        3}$ A.\,W.\ Blain,$^{\! 4}$ J.-P.\ Kneib,$^{\! 3}$ \and
        J.\ B\'ezecourt,$^{\! 3}$ T.\,H.\ Kerr$^5$  and J.\,K.\ Davies$^5$
        \vspace*{1mm}\\
        $^1$ Institute for Astronomy, Dept.\ of Physics \& Astronomy, 
        University of Edinburgh, Blackford Hill, Edinburgh EH9 3HJ\\
        $^2$ Department of Physics, University of Durham, South Road,
        Durham DH1 3LE\\
        $^3$ Observatoire Midi-Pyr\'en\'ees, 14 Avenue E.\ Belin,
        F-31400 Toulouse, France\\
        $^4$ Cavendish Laboratory, Madingley Road, Cambridge CB3 0HE\\
        $^5$ Joint Astronomy Centre, 660 N.\ A'oh\=ok\=u Place, University
        Park, Hilo HI 96720, USA}

\date{\fbox{\sc Accepted}}

\pagerange{000--000}

\begin{document}

\maketitle

\begin{abstract}
We present a detailed study of SMM\,02399$-$0136, a hyperluminous,
active galaxy selected from a sub-mm survey of the distant Universe.
This galaxy is the brightest source in the fields of seven rich,
lensing clusters, with a total area of 0.01\,degree$^2$, that we have
mapped with a sensitivity of $\sim$2\,mJy\,beam$^{-1}$ at 850\,$\mu$m.
We identify a compact optical counterpart with an apparent magnitude
of $B\sim23$ and a low-surface-brightness companion $\sim$3\,arcsec
away. Our spectroscopy shows that both components have the same
redshift; $z = 2.803 \pm 0.003$.  The emission line widths,
FWHM\,$\simeq $\,1000--1500\,km\,s$^{-1}$, and line ratios, along with
the compact morphology and high luminosity ($M_B \simeq -24.0$) of the
galaxy indicate that SMM\,02399$-$0136 contains a rare dust-embedded,
narrow-line or type-2 active galactic nucleus (AGN).  The source is
gravitationally lensed by the foreground cluster, amplifying its
apparent luminosity by a factor of 2.5, and our detailed lens model
allows us to accurately correct for this.  Taking the amplification
into account we estimate that SMM\,02399$-$0136 is intrinsically a
factor of five times more luminous than {\em IRAS} F\,10214+4724. Its
far-infrared and H$\alpha$ luminosities and low-surface-brightness
radio emission are indicative of an interaction-induced starburst and
the star-formation rate (SFR) could be several thousand solar masses
per year. This, however, assumes that the starburst is the dominant
source of energy, but we cannot yet determine reliably the relative
contributions of the starburst and the buried AGN. A dust mass of
5--7$\times 10^8$\,M$_{\odot}$ is indicated by our data for a dust
temperature of 40--50\,{\sc k}, independent of the dominant energy
source. We estimate the possible space density of such luminous sub-mm
sources, and find that while a large population of these obscured
sources could be detected in future wide-field sub-mm surveys, they
are unlikely to dominate the faint counts in this waveband. Galaxies
such as SMM\,02399$-$0136 and F\,10214+4724 cannot be easily detected
in conventional AGN/QSO surveys, and so estimates of the prevalence of
AGN in the early Universe may require significant revision.
\end{abstract}


\section{Introduction}

There is abundant evidence that a large amount of dust exists in
galaxies in the distant Universe. Many sub-mm detections of known
high-redshift radio galaxies and quasars have been made following the
identification of the ultraluminous {\it IRAS} source, F\,10214+4724,
at $z=2.286$ (Rowan-Robinson et al.\ 1991), for example: 4C\,41.17
(Dunlop et al.\ 1994), BR\,1202$-$0725 (Isaak et al.\ 1994) and
8C\,1435+635 (Ivison et al.\ 1998).  These observations detect
starlight, and/or quasar light, re-radiated by dust in the rest-frame
far-IR waveband. The discovery of F\,10214+4724 suggested that the
{\em selection} of distant, star-forming sources in the sub-mm would
become commonplace.  This is still the perceived wisdom, although more
recent work has provided a cautionary lesson by demonstrating that the
extremely high luminosity of F\,10214+4724 is due to a
strongly-lensed, dust-embedded AGN rather than solely to a starburst
(e.g.\ Broadhurst \& Lehar 1995; Eisenhardt et al.\ 1996; Goodrich et
al.\ 1996; Serjeant et al.\ 1998).

Nevertheless, the sub-mm waveband still offers excellent prospects for
the detection of a large population of very distant, star-forming
galaxies and, based on F\,10214+4724, dusty AGN. The steep slope of
the modified blackbody spectrum of dusty galaxies in the sub-mm and
far-IR wavebands leads to large, negative {\em K}--corrections, which
can even be sufficient to overcome the effects of the inverse square
law (Blain \& Longair 1993). A flat flux density--redshift relation is
predicted for star-forming galaxies and dusty AGN at $z > 0.5$, and so
the selection function of a sub-mm survey is expected to be relatively
constant and to extend out to redshifts $z \sim 10$.  This unusual
selection function is expected to increase the detectability of lensed
galaxies in sub-mm surveys (Blain 1996a, 1997a,b).

The first discovery of a population of galaxies in the sub-mm waveband
has recently been made with the new Sub-mm Common-User Bolometer Array
(SCUBA --- Holland et al.\ 1998) on the 15-m James Clerk Maxwell
Telescope (JCMT).  Smail, Ivison \& Blain (1997; SIB) took advantage
of the gravitational amplification of the background sky seen through
rich, lensing clusters to undertake the first, effectively
blank-field, sub-mm survey.  The sources detected in the fields of the
rich cluster lenses, A\,370 and Cl\,2244$-$02, have allowed the sub-mm
counts of distant galaxies to be measured for the first time, as
compared with recent upper limits reported at other mm and sub-mm
wavelengths by Phillips (1997) and Wilner \& Wright (1997). Our
dataset has now expanded to cover seven independent fields (Blain et
al.\ 1998), all mapped with a sensitivity of $\sim$2\,mJy\,beam$^{-1}$
at 850\,$\mu$m. The cumulative 850-$\mu$m source counts at faint flux
densities are substantially higher than those predicted by a simple
non-evolving extrapolation of the local 60-$\mu$m galaxy luminosity
function (Saunders et al.\ 1990). SIB concluded that the comoving
star-formation density in the Universe associated with sources that
have apparently very high SFRs, in excess of about
100\,M$_{\odot}$\,yr$^{-1}$, evolves strongly between $z=0$ and $z >
1$ (also see Blain et al.\ 1998).

In view of the impact of the sub-mm surveys on our view of galaxy
formation in the distant Universe, a more detailed study of the
individual galaxies detected by SIB is clearly warranted. Here we
present the results of follow-up observations of the brightest of
these sub-mm sources, SMM\,02399$-$0136, in the field of A\,370.  A
similar analysis of our full sub-mm-selected sample of galaxies is
presented in Smail et al.\ (1998).  We have exploited the
excellent archival data available for these rich clusters, as well as
obtaining new data, in order to investigate the properties of
SMM\,02399$-$0136.  First, we discuss the archival data; deep
$U\!BV\!RI$ images, ultra-deep 1.4- and 8.7-GHz radio maps and limits
from {\em ROSAT} and {\em IRAS}. Secondly, we introduce the new
observations; more sensitive SCUBA sub-mm maps and photometry,
CFHT/UKIRT optical/near-IR imaging and spectroscopy and CAM imaging
data from {\em ISO}.  Thirdly, we discuss the nature of the source,
and finally its implications for the interpretation of our sub-mm
survey and the population of AGN in the distant universe.  Throughout
we assume $\Omega_0=1$ and $H_0 = 50$\,km\,s$^{-1}$\,Mpc$^{-1}$.

\section{Observations}

SMM\,02399$-$0136 is the brightest sub-mm source detected in a survey
which covers 36\,arcmin$^2$ (Blain et al.\ 1998; Smail et al.\ 1998).
It lies in the field of the $z=0.37$ cluster, A\,370, and is detected
both in the 850-$\mu$m map at high significance (also appearing in the
reference beams) and at a lower signal-to-noise ratio at 450\,$\mu$m
(Fig.\,1).  The original maps (SIB) have now been augmented by a
further 10-ks integration, giving a total integration time of 33.7\,ks
and an rms noise level of $\sim$1.7\,mJy\,beam$^{-1}$. The 33.7-ks
maps can be used to determine a position $\alpha = 02^{\rm h} 39^{\rm
m} 51.96^{\rm s}$, $\delta = -01^{\circ} 35' 59.0''$ (J2000) that is
accurate to within 3\,arcsec. All the observations are summarized in
Table~1.

%
%
\begin{table*}
\begin{center}
\caption{ \hfil The observed properties of SMM\,02399$-$0136 from the radio 
to the UV waveband$^{\star}$.\hfil }
\begin{tabular}{lcccl}
\noalign{\medskip}
\noalign{\smallskip}
Property & Telescope & \multispan2{ \hbox{SMM\,02399$-$0136} }& {Comment} \cr
           &  & L1         &       L2        & \cr
\noalign{\medskip}
$\alpha$(J2000) & & \hbox{$02^{\rm h} 39^{\rm m} 51.88^{\rm s}$}& \hbox{$02^{\rm h} 39^{\rm m} 52.10^{\rm s}$} & Position from $I$ image, accurate to $\pm 1$\,arcsec. \cr
$\delta$(J2000) & & \hbox{$-01^{\circ} 35' 58.0''$}& \hbox{$-01^{\circ} 35' 57.2''$} & Position from $I$ image, accurate to $\pm 1$\,arcsec.\cr
\noalign{\medskip}
Redshift  & & \hbox{$2.803\pm 0.003$}& \hbox{$2.799\pm 0.003$} & See Table~2. \cr
\noalign{\medskip}
Flux density at: & & & & \cr
\noalign{\smallskip}
$\>\>350\,\mu$m & JCMT & \multispan2{ \hbox{$<$323\,mJy$^{\dagger}$} }& 3$\sigma$ limit. Photometry mode. \cr
$\>\>450\,\mu$m & JCMT & \multispan2{ \hbox{$69 \pm 15$\,mJy$^{\dagger}$} }& Using new, deeper map (photometry mode: $52\pm19$\,mJy). \cr
$\>\>750\,\mu$m & JCMT & \multispan2{ \hbox{$28 \pm 5$\,mJy$^{\dagger}$} }& Photometry mode. \cr
$\>\>850\,\mu$m & JCMT & \multispan2{ \hbox{$26 \pm 3$\,mJy$^{\dagger}$} }& Using new, deeper map (photometry mode: $24\pm3$\,mJy). \cr
$\>\>1350\,\mu$m & JCMT & \multispan2{ \hbox{$5.7 \pm 1.0$\,mJy$^{\dagger}$} }&  Photometry mode.\cr
$\>\>2000\,\mu$m & JCMT & \multispan2{ \hbox{$<$8.4\,mJy$^{\dagger}$} }& 3$\sigma$ limit. Photometry mode.\cr
\noalign{\smallskip}
$\>\>12\,\mu$m & {\em IRAS} & \multispan2{ \hbox{$<$91\,mJy} }& Estimate from nearby FSC sources. \cr
$\>\>25\,\mu$m & {\em IRAS} & \multispan2{ \hbox{$<$86\,mJy} }& Estimate from nearby FSC sources. \cr
$\>\>60\,\mu$m & {\em IRAS} & \multispan2{ \hbox{$<$428\,mJy} }& Estimate from nearby FSC sources.\cr
$\>\>100\,\mu$m & {\em IRAS} & \multispan2{ \hbox{$<$715\,mJy} }& Estimate from nearby FSC sources. \cr
\noalign{\smallskip}
$\>\>15\,\mu$m & {\em ISO}/CAM & \multispan2{ \hbox{$1.2 \pm 0.4$\,mJy$^{\dagger}$} }& Metcalfe et al.\ (1998). \cr
\noalign{\smallskip}
$\>\>3.4$\,cm & VLA (D)& \multispan2{ \hbox{$<$57\,$\mu$Jy} }&  3$\sigma$ limit, 8.7\,GHz. \cr
$\>\>21.5$\,cm & VLA (A)& \multispan2{ \hbox{$526 \pm 50$\,$\mu$Jy} }& 1.4\,GHz. \cr
\noalign{\bigskip}
$B_{\rm tot}$ & CFHT & \hbox{$22.69\pm 0.03$} & \hbox{$23.73\pm 0.05$} & Total magnitude \cr
$U_{\rm ap}$  & {\em HST} & \hbox{$25.17\pm 0.19$} & \hbox{$> 25.4$} & Aperture magnitude, 3$\sigma$ limit for L2. \cr
$B_{\rm ap}$ & CFHT & \hbox{$23.88\pm 0.04$} & \hbox{$25.26\pm 0.09$} &  Aperture magnitude \cr
$V_{\rm ap}$ & Danish 1.5-m & \hbox{$24.01\pm 0.04$} & \hbox{$25.62\pm 0.11$} &  Aperture magnitude \cr
$R_{\rm ap}$ & CFHT & \hbox{$22.60\pm 0.04$} & \hbox{$24.43\pm 0.11$} &  Aperture magnitude \cr
$I_{\rm ap}$ & CFHT & \hbox{$21.88\pm 0.05$} & \hbox{$23.92\pm 0.15$} &  Aperture magnitude \cr
$K_{\rm ap}$ & UKIRT & \hbox{$19.10\pm 0.11$} & \hbox{$21.83\pm 0.73$} & Aperture magnitude \cr
\noalign{\medskip}
{0.1--2.0\,keV} & {\em ROSAT}/HRI & \multispan2{ \hbox{$< 10^{-14}$ erg\,s$^{-1}$\,cm$^{-2}$} }&  3$\sigma$ limit. \cr
\end{tabular}
\end{center}

\noindent
$^{\star}$ None of these quantities have been corrected for the
effects of gravitational lensing.

\noindent
$^{\dagger}$ In a break from sub-mm/far-IR convention, errors include the 
uncertainty in calibration.
\end{table*}
\bigskip

\subsection{Archival data}

We targeted distant, massive clusters in our sub-mm observations
primarily to exploit the gravitational lensing amplification of
background sources.  However, these targets are also richly endowed
with archival data of excellent quality, covering the whole spectrum
from the radio to the X-ray, and providing a wealth of additional
information on our sub-mm detections.

\subsubsection{VLA observations}

A\,370 has been observed at radio wavelengths, both to
understand the radio properties of member galaxies and in the hope of
identifying lensed radio sources seen through the cluster core.  We
have used these observations to place limits on the intensity of the
radio emission from SMM\,02399$-$0136 and to better define its
position.

The most useful radio observation for our purpose is a very deep
1.4-GHz A-configuration Very Large Array (VLA) map (Owen \&
Dwarakanath 1998, in preparation; Fig.\,1), with a noise level of only
10\,$\mu$Jy\,beam$^{-1}$. This image clearly shows a weak, slightly
extended source at $\alpha = 02^{\rm h} 39^{\rm m} 51.82^{\rm s}$,
$\delta = -01^{\circ} 36' 00.6''$ (J2000), within the sub-mm
positional uncertainty of SMM\,02399$-$0136, and within 1\,arcsec of a
compact optical source (\S2.1.2). After deconvolution with the
synthesized beam, the emission region is found to be 7.9\,arcsec by
2.2\,arcsec in size, with a position angle of 71$^{\circ}$, a maximum
surface brightness of 221\,$\mu$Jy\,beam$^{-1}$ and an integrated flux
density of $526 \pm 50$\,$\mu$Jy, just below the detection thresholds
of radio surveys such as the Leiden--Berkeley Deep Survey (Windhorst,
van Heerde \& Katgert 1984) which has a selection limit of 0.6\,mJy at
1.4\,GHz. The 1.4-GHz flux density corresponds to a rest-frame 5.3-GHz
power of about $2.6 \times 10^{23}$\,W\,Hz$^{-1}$\,sr$^{-1}$.  At
higher frequencies, H.\ Liang has kindly inspected a deep 8.7-GHz
D-configuration VLA continuum map (Liang 1997; Fig.\,1) with a
resolution of about 8\,arcsec, and placed a 3$\sigma$ limit of
57\,$\mu$Jy to the 8.7-GHz flux density.

\subsubsection{Optical imaging}

A\,370 has been studied extensively in the optical since the first
giant gravitational arc was identified in its core by Soucail et al.\
(1987).  Hence, we have been able to gather together a number of
high-quality images.  These include very deep $BRI$ images taken using
the 3.6-m Canada--France--Hawaii Telescope (CFHT) in superb
conditions, with 0.4-arcsec seeing in the $I$-band and 0.6-arcsec
seeing in $B$ and $R$; total integration times were 17.5\,ks in $B$
and 9.0\,ks in both $R$ and $I$. More information is given in Kneib et
al.\ (1993, 1994). We also have a 18.0-ks integration in $V$ with
1.3-arcsec seeing from the Danish 1.5-m telescope at La Silla, Chile
(Smail et al.\ 1991).

A\,370 has also been observed by the {\em Hubble Space Telescope}
({\em HST}\,) but, unfortunately, SMM\,02399$-$0136 falls outside the
field of view of the post-refurbishment WFPC2 observations by Saglia
(GO\,6003), and while the source is included in the pre-refurbishment
WF/PC1 F702W ($R$-band) image of Dressler (GO\,2373), discussed by
Couch et al.\ (1994), this image lacks the surface brightness
sensitivity necessary to provide detailed morphological
information. SMM\,02399$-$0136 was also included in a 16.4-ks WFPC2
F336W ($U$-band) image of Deharveng (GO\,5709), which samples the
rest-frame spectral energy distribution (SED) of the source at
wavelengths between 830 and 930\,\AA. For aperture photometry with
this data, we adopt the zero-point from Table\,7 of Holtzman et al.\
(1995), and assume $(U-B)=0$ to convert between the F336W and $U$-band
magnitudes.

%
%
\begin{figure*}
\centerline{\psfig{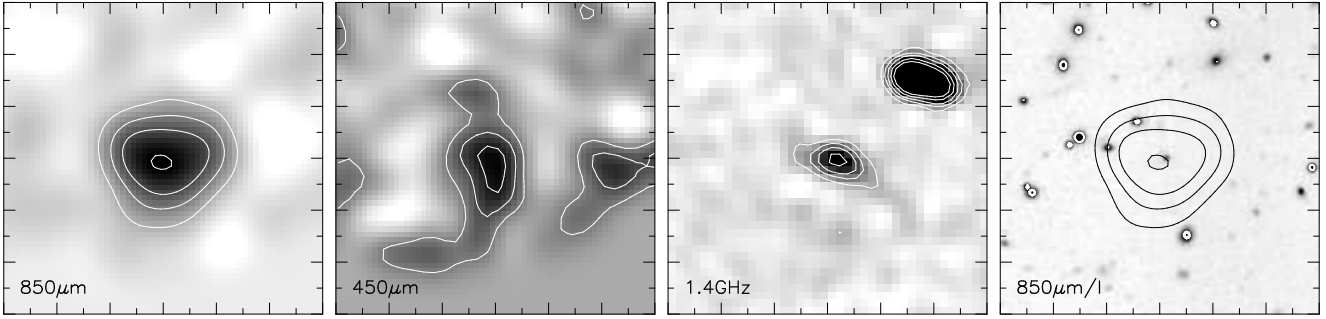}}
\caption{ Maps of SMM\,02399$-$0136. From left to right: 850-$\mu$m
SCUBA image; 450-$\mu$m SCUBA image; 1.4-GHz VLA A-configuration map;
and the 850-$\mu$m SCUBA map overlaid on a CFHT $I$-band image.  North
is up and East is to the left. Each panel is 1\,arcmin square.}
\end{figure*}
%
%
\begin{figure*}
\centerline{\psfig{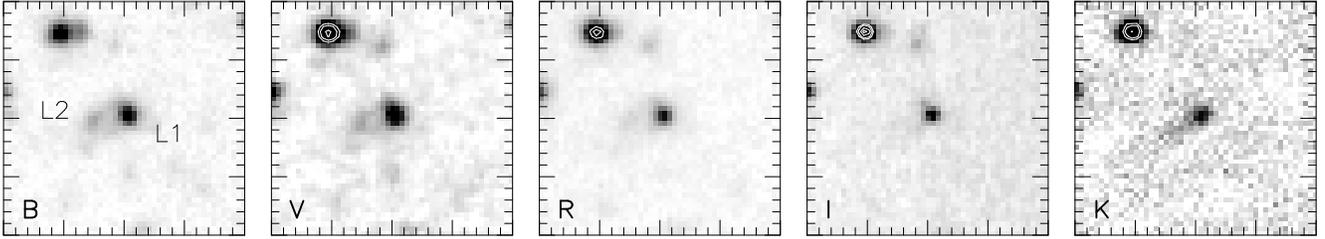}}
\caption{ $BV\!RI\!K$ images of SMM\,02399$-$0136.  The two optical
counterparts, L1 and L2, are identified in the $B$-band exposure; both
have red colours across these passbands.  Note that L1 appears very
compact in all passbands, even at the very fine 0.4--0.6-arcsec
resolution of the CFHT $BRI$ exposures. The low-surface-brightness
feature L2 may be either a companion or the remnant of a tidal
interaction that triggered the activity in L1. North is up and East is
to the left. Each panel is 20\,arcsec square.}
\end{figure*}

We find two faint optical sources within 3\,arcsec of the 1.4-GHz
position (\S2.1.1): SMM\,02399$-$0136[L1] and [L2] (Figs\,1 \& 2). L1
lies within 1\,arcsec of the radio position, while L2 lies 3\,arcsec
to the East, at a position angle of about 88$^{\circ}$. Two other
considerably fainter sources are visible in the $I$-band image to the
N and NNE within 2\,arcsec of L1; however, because of their faintness
and the coarser angular resolution of our other optical images, these
features are not visible in the other bands, and so we concentrate on
L1 and L2 below. The results of aperture photometry in 1.5-arcsec
diameter apertures from the seeing-matched $U\!BV\!RI\!K$ images are
listed in Table\,1, along with estimates of the total $B$-band
magnitude of each component. The reddening towards A\,370 is
$E(B-V)=0.015$, and so no foreground reddening correction has been
applied to the results in Table\,1.

L1 is brighter than L2 in our high-resolution $I$-band image (Fig.\,2;
Table\,1). It appears to have a compact morphology, and is marginally
resolved with an intrinsic FWHM of 0.3\,arcsec. The WF/PC1 F702W and
WFPC2 F336W images show an unresolved source, with a FWHM of about
0.1--0.2\,arcsec, coincident with L1. The F336W exposure also shows
some emission associated with L2. From our ground-based images we can
see that L2 has a diffuse morphology with a profile which extends out
to L1 (Fig.\,2). Source morphologies are discussed in more detail in
\S3.2.

\subsubsection{Observations in other wavebands}

A deep {\em ROSAT} High Resolution Imager exposure of A\,370 with a
total exposure time of 30.8\,ks, taken by B\"ohringer, was retrieved
from the {\em ROSAT} HEASARC Archive at GSFC. The positional accuracy
is good to within about 8\,arcsec, based on the relative positions of
bright stars and galaxies in this exposure and our wide-field optical
images. Due to the extended emission from the hot intracluster gas in
the foreground cluster, it is difficult to identify any obvious
emission from SMM\,02399$-$0136. By fitting and subtracting a smooth
background we estimate a 3$\sigma$ upper limit of $S_\nu \ls 10^{-14}$
erg\,s$^{-1}$\,cm$^{-2}$ to the flux in the 0.1--2\,keV band from a
point source at this position in a 24-arcsec aperture. Assuming $S_\nu
\propto \nu^{+\alpha}$, with $\alpha\simeq -0.5$ for an obscured
source (Almaini et al.\ 1995), and a Galactic H${\sc i}$ column
density of about $3\times 10^{20}$\,cm$^{-2}$, this flux density
corresponds to a luminosity of less than $5 \times
10^{44}$\,erg\,s$^{-1}$ in the 2--10\,keV band at $z=2.8$ (\S2.2.2),
after correcting for lens amplification, which is consistent with a
type-2 AGN (c.f.\ Lawrence et al.\ 1994).

The relatively uninteresting {\em IRAS} All-Sky Survey upper limits at
12, 25, 60 and 100\,$\mu$m are listed in Table~1. {\em Infrared Space
Observatory} ({\em ISO}) continuum observations at 175\,$\mu$m are
discussed by Frayer et al.\ (1998, in preparation), together with
observations of the molecular gas associated with SMM\,02399$-$0136.

\subsection{New targeted observations}

\subsubsection{New measurements with SCUBA on the JCMT}

Follow-up observations with SCUBA have been made using the photometry
mode at 0.35, 0.45, 0.75, 0.85, 1.35 and 2.0\,mm. As reported by SIB,
the secondary mirror was chopped at 6.944\,Hz, but instead of the
previous, complex jiggle pattern, a simple $3 \times 3$ jiggle was
performed (with 2-arcsec offsets) in both the signal and reference
beams, nodding between the two every 9\,s in a
signal--reference--reference--signal pattern. Every hour the pointing
was checked on the blazar, 0336$-$019, and a skydip was performed to
measure the atmospheric opacity. The rms pointing errors were around
2\,arcsec.

The data were reduced by subtracting the measurements in the reference
beam from those in the signal beam, rejecting obvious spikes. Where
possible, residual sky background emission was removed using the
technique described by Jenness, Lightfoot \& Holland (1998). The data
were then corrected for atmospheric opacity and calibrated against
Uranus. The measured flux densities, and those derived from the deeper
450- and 850-$\mu$m maps discussed earlier in \S2, are listed in
Table~1.

\subsubsection{Optical spectroscopy from CFHT}

%
%
\begin{figure*}
\centerline{\psfig{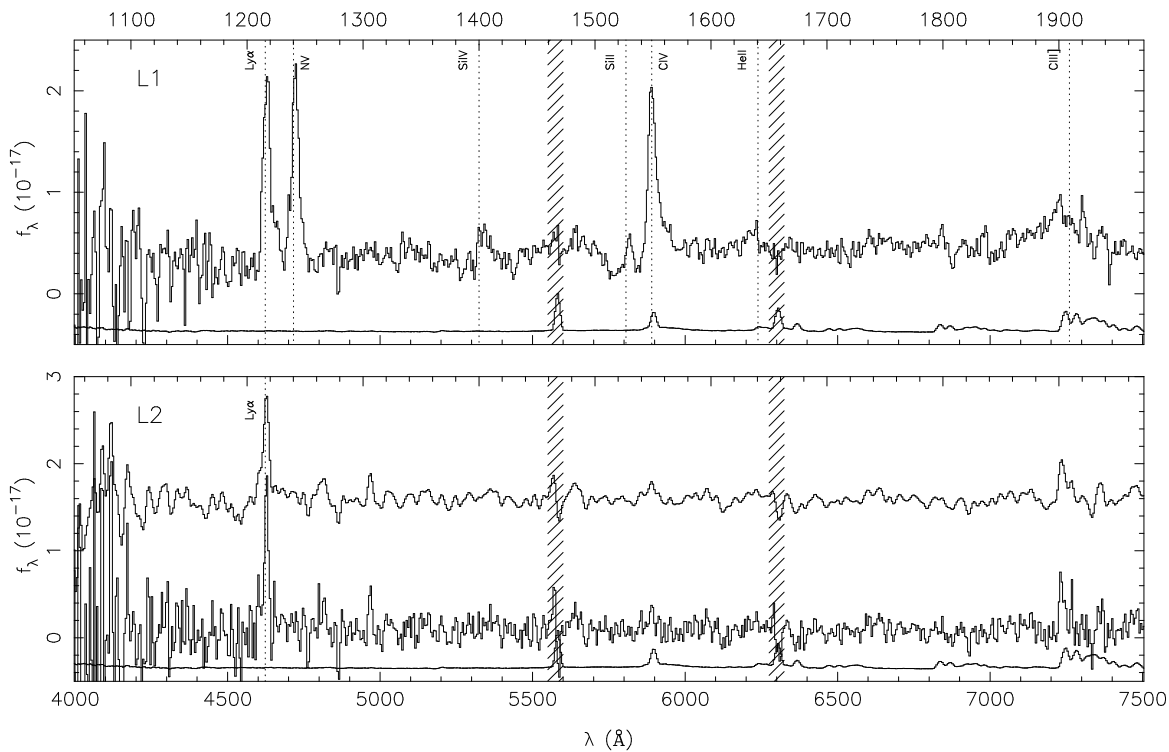}}
\caption{Optical spectra of both L1 and L2 from the MOS spectrograph
on CFHT. The spectrum of L1 shows a number of narrow emission lines at
$z=2.803$ superimposed on a blue continuum, with hints of broad
absorption features on the blue wings of some lines. The spectrum of
the substantially fainter L2 (shown both raw and at the instrumental
resolution in the lower panel) shows only Ly$\alpha$ at a similar
redshift to that of L1. The lower spectrum in each panel is an
arbitrarily scaled sky spectrum, which indicates the positions of sky
emission features; the hatched regions are strongly affected. The
identified optical lines listed in Table\,2 are indicated. The
observed wavelength is shown on the bottom axis: the rest-frame
wavelength at $z=2.8$ is shown on the top axis. The spectra of L1 and
L2 are flux calibrated and are plotted in units of
$10^{-17}$\,erg\,cm$^{-2}$\,s$^{-1}$\,\AA$^{-1}$.}
\end{figure*}
%
%
\begin{figure}
\centerline{\psfig{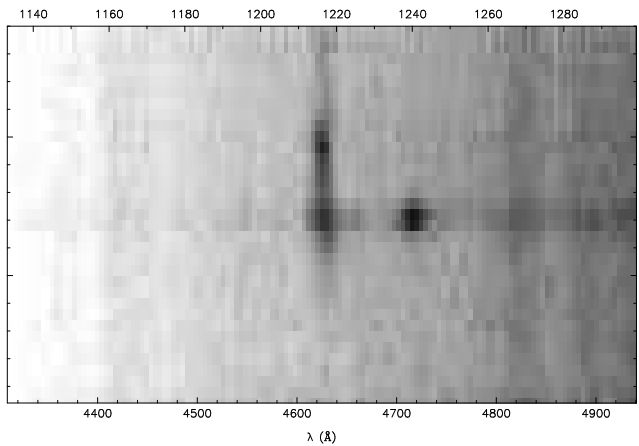}}
\caption{The two-dimensional optical spectrum of L1/L2 around the
wavelength of the redshifted Ly$\alpha$ line.  L1 is the lower of the
two galaxies, a compact galaxy with strong Ly$\alpha$ and N{\sc v}
emission lines; in contrast, L2 is clearly extended and shows only
narrow Ly$\alpha$ emission.  The observed and rest-frame wavelengths
are shown on the bottom and top axes respectively; the tick
marks on the vertical axis are separated by 1\,arcsec.}
\end{figure}

Initial spectroscopy of SMM\,02399$-$0136 was undertaken with the MOS
multi-object spectrograph on the 3.6-m CFHT, Mauna Kea, on the nights
of 1997 Sept 5--7.  The $2{\rm k}^2$ SITe CCD was used with the O300
grism and slitlets 15\,arcsec in length and 1.2\,arcsec in width,
which were aligned along the line joining L1 and L2 (a position angle
of $88.6^{\circ}$). A number of both sub-mm- and optically-selected
galaxies were targeted across the field of A\,370. The mask containing
SMM\,02399$-$0136 was observed for a total of 24.3\,ks, in nine 2.7-ks
exposures. The data were debiased and flatfielded using dome flats in
the standard manner, and then the one-dimensional spectra of both L1
and L2 were optimally extracted.  The spectra were calibrated in
wavelength using He--Ar arc spectra that were interspersed between the
science exposures, and in flux density using observations of
Feige\,110 (Oke 1990). The final spectra (Fig.\,3) cover the
wavelength range 3500--10000\,\AA\ at high sensitivity, with
4.7-\AA\,pixel$^{-1}$ sampling and an effective resolution of about
15\,\AA\ at 4500\,\AA; the wavelength calibration is accurate to
within 0.1\,\AA\ rms. At wavelengths longer than 7500\,\AA\ the data
are contaminated by the second-order spectra, and so these wavelengths
are not shown in Fig.\,3.  To demonstrate the spatial separation of L1
and L2, and to highlight their different emission characteristics, a
section of the two-dimensional spectrum around the Ly$\alpha$ line is
shown in Fig.\,4.

A number of narrow emission lines and absorption features superimposed
on a blue continuum can be identified in the spectrum of L1 (Table~2),
the most prominent of these lines are redshifted
Ly$\alpha$\,1215.7\AA, N\,{\sc v}\,1240.2\AA\ and C\,{\sc
iv}\,1549.0\AA. Using all the features listed in Table\,2, we estimate
a redshift $z=2.803\pm 0.003$, at which an angular scale of 1\,arcsec
corresponds to 7.45\,kpc. The FWHM of these emission lines are
typically about 26\AA, which is equivalent to about
1000--1500\,km\,s$^{-1}$ in the rest frame after correcting for the
instrumental broadening (Table\,2). We see no evidence for a broad
component of these lines, although some are slightly asymmetric. This
asymmetry may be related to blue absorption troughs, the most
prominent of which are associated with C\,{\sc iv}/Si\,{\sc ii} and
Si\,{\sc iv}.  The typical width of these troughs is about 70--150\AA,
corresponding to about 3000--7000\,km\,s$^{-1}$ at $z=2.8$.  We do see
one broad line in the spectrum of L1 at 7000--7400\AA, although this
lies on an atmospheric OH bandhead.  We identify this feature as broad
C\,{\sc iii}]\,1909\AA, with an apparent FWHM of about 180\AA, or
around 7400\,km\,s$^{-1}$ in the rest frame (c.f.\ Serjeant et al.\
1998; although see Hartig \& Baldwin 1986 for other possible
identifications).  The only feature of interest outside the wavelength
range shown in Fig.\,3 is a possible detection of a very weak
Ly$\beta$ line at a rest-frame wavelength of 1026\AA.  These features
are discussed in more detail in \S3.3.

The only strong line detected in the spectrum of the fainter, diffuse
structure, L2, is Ly$\alpha$\,1215.7\AA\ at a similar wavelength to that
observed in L1; there is little or no continuum (Figs\,3 \& 4). The
deconvolved velocity width of the Ly$\alpha$ line is about
500\,km\,s$^{-1}$, much narrower than the equivalent line in L1
(Table\,2).  Looking at Fig.\,4 we see that the Ly$\alpha$ emission
extends beyond the peak in L2 out to a distance of at least 8\,arcsec
from L1, beyond the continuum extent of the source shown in Fig.\,2.

%
%
\begin{table*}
\begin{center}
\caption{ \hfil Spectral-line identifications in SMM\,02399$-$0136
[L1] and [L2]. $\lambda_{\rm obs}$ is the observed wavelength,
$\lambda_{0}$ is the emitted wavelength if the redshift of the line is
$z$.  \hfil }
\begin{tabular}{lcccrrl}
\noalign{\medskip}
\noalign{\smallskip}
{Line} & {$\lambda_{\rm obs}$}& {$\lambda_{0}$} & {$\>\>z\>\>$} & 
{EW} & {FWHM} & {Comments} \cr
& (\AA) & (\AA) &  & (\AA) & (km\,s$^{-1}$) & \cr
\noalign{\medskip}
{\bf L1} & & & & & \cr
Ly\,$\alpha$ & 4623.90 & 1215.7 & 2.804 & $-195.3$ & 1850~~~ & Red wing. \cr
N\,{\sc v} & 4717.53 & 1240.2 & 2.804 & $-167.1$ & 1790~~~ & Red wing. \cr
Si\,{\sc iv} + O\,{\sc iv}] & 5335.00 & $\sim$1400 & $\simeq 2.8$ 
& $-23.4$ & 1790~~~ & Absorption trough to blue. \cr
Si\,{\sc ii} & 5813.05 & 1526.7 & 2.808 & $-33.7$ & 1160~~~ & Possible absorption 
trough to blue, \cr 
& & & & & & lies in C{\sc iv} trough. \cr
C\,{\sc iv} & 5887.76 & 1549.0 & 2.801 & $-111.7$ & 1560~~~ & Red wing, absorption trough to blue. \cr
He\,{\sc ii} & 6222.19 & 1640.5 & 2.793 & $-19.6$ & 2800~~~ & \cr
C\,{\sc iii}] & 7220.84 & 1908.7 & 2.783 & $-156.2$ & 7700~~~ & Very broad feature on OH band head.\cr
\noalign{\smallskip}
[O\,{\sc iii}] & 18991 & 5006.9 & $\simeq 2.8$  
& ... & 1020~~~ & 
$\sim 8\times 10^{-18}$ W\,m$^{-2}$. \cr
H\,$\alpha$/[N\,{\sc ii}] & $\simeq 24881$ 
& 6548/6563 & $\simeq 2.8$  
& ... & 1060~~~ & $\sim 9\times 10^{-18}$ W\,m$^{-2}$. \cr
\noalign{\medskip}
{\bf L2} & & & & & \cr
Ly\,$\alpha$ & 4622.96 & 1215.7 & 2.803 & $-250$~ &  900~~~ &  \cr
\noalign{\smallskip}
[O\,{\sc iii}] & 19050 & 5006.9 & 2.805 & ... & 1600~~~ & $\sim 2\times 10^{-17}$ W\,m$^{-2}$. \cr
H\,$\alpha$/[N{\sc ii}] & $\simeq 24905$ 
& 6548/6563 & $\simeq 2.8$  
& ... & 680~~~ & $\sim 5\times 10^{-18}$ W\,m$^{-2}$. \cr
\end{tabular}
\end{center}
\end{table*}

\subsubsection{Infrared imaging and spectroscopy with UKIRT}

During 1997 July 18, we obtained a $K$-band image of the A\,370 field
using the IRCAM3 near-IR camera on the 3.8-m UK Infrared Telescope
(UKIRT) in 1.0-arcsec seeing (T.\,Naylor, private communication). We
built up a total integration time of 4.3\,ks using the technique
described by Dunlop \& Peacock (1993). The resulting image is shown in
Fig.\,2: L1 and an extension in the direction of L2 are clearly
visible. Photometry from this image is listed in Table~1.

A $K$-band spectrum was obtained using the CGS4 spectrometer on UKIRT
during 1997 Oct 5 in good conditions and 0.7-arcsec seeing.  The
40-lines\,mm$^{-1}$ grating was used to cover the entire $K$-band
window at a resolution of 560\,km\,s$^{-1}$. A 90-arcsec-long,
1.22-arcsec-wide slit was used at a position angle of $88^{\circ}$,
the data were Nyquist sampled and the telescope was nodded
18.3\,arcsec along the slit every 80-s to allow adequate sky
subtraction. Offsets from a nearby bright star were used to position
the slit accurately on the galaxy. The total exposure time was
4.6-ks. Flux calibration and telluric line cancellation were
performed by ratioing the spectrum with that of CMC\,100796, a F0V
star, and then multiplying by a blackbody spectrum appropriate to the
temperature and magnitude of that star.

Continuum emission can be discerned in the spectrum of L1. The
[O\,{\sc iii}] 4959 and 5007\AA\ lines are present at a 4$\sigma$
significance, blended together at the blue end of the spectrum,
corresponding to $z\simeq 2.793$. They are blueshifted slightly with
respect to the optical emission line, although there are no arc lines
near this wavelength, and so the uncertainty in the wavelength
calibration is increased. The flux density in the [O\,{\sc iii}]
5007\AA\ line is about $8\times 10^{-18}$\,W\,m$^{-2}$. Another line
is also present at the extreme red end of the spectrum at 3$\sigma$
significance. Its wavelength is consistent with either pure H$\alpha$
6563\AA\, or a blend of H$\alpha$ and [N\,{\sc ii}] 6548\AA.  The flux
density in the 1060\,km\,s$^{-1}$ wide line is about $9\times
10^{-18}$\,W\,m$^{-2}$. If it is H$\alpha$, then the redshift is $z =
2.795\pm0.009$.

The [O\,{\sc iii}] 5007\AA\ line is detected in the spectrum of L2 at
5$\sigma$ significance, with [O\,{\sc iii}] 4959\AA\ blended into its
blue wing.  A twin Gaussian profile, in which one component is three
times brighter than the other but with an identical width, fits the
profile very accurately at the expected wavelength ratio and $z\simeq
2.805$, although again the wavelength calibration may be slightly
uncertain. The H$\alpha$ or H$\alpha$/[N\,{\sc ii}] blend seen towards
L1 is also present in the spectrum of L2, but only at the 2$\sigma$
level, corresponding to a flux density of about $5\times
10^{-18}$\,W\,m$^{-2}$.

\subsubsection{Mid-infrared data from ISO}

During 1996 Aug 18--19, Metcalfe et al.\ (1998) obtained an extremely
deep mid-IR image of A\,370, targeting the giant arc, A0, with the
long-wavelength channel of the 32$^2$-pixel camera ISOCAM (C\'esarsky
et al.\ 1996) using a 6-$\mu$m-wide filter centred at a wavelength of
15\,$\mu$m. The total integration time near SMM\,02399$-$0136 was
1.6\,ks, and the galaxy was detected at a significance of about
40$\sigma$, although not resolved. The 15-$\mu$m flux density is $1.2
\pm 0.4$\,mJy after accounting for the uncertainty in calibration
(L.\,Metcalfe, private communication).

\section{Results}

\subsection{Gravitational lensing}

Before discussing the detailed continuum and spectral-line properties
of SMM\,02399$-$0136, it is important to clarify the degree of
amplification experienced due to the massive, concentrated, foreground
cluster.  Detailed mass models (Kneib et al.\ 1993; B\'ezecourt et
al.\ 1998, in preparation) suggest that for a $z=2.8$ galaxy at this
position in the image plane the most likely amplification factor is
2.5, with a robust upper limit of 5. A factor of 2.5 is assumed below,
and this correction is applied to all the derived physical quantities
listed below, for example $M_B$ and dust mass.

\subsection{Morphology}

The UV and optical images (Fig.\,2) indicate that SMM\,02399$-$0136
has two related counterparts, L1 and L2.  L1, the compact component,
may be marginally resolved with an intrinsic FWHM of 0.3\,arcsec,
corresponding to about 1\,kpc in our adopted cosmology at $z=2.8$
after correcting for lens amplification.  The angular separation of L1
and the brightness peak in L2 is about 3\,arcsec, corresponding to
22\,kpc, or $\sim 9$\,kpc after correcting for amplification in the
tangential direction, this is roughly along the line joining L1 and
L2. L2 has a more complex morphology than L1, showing a ridge of
emission to the North and a relatively diffuse region extending South
and West towards L1.

The radio emission from the 1.4-GHz maps is also resolved in the
East-West direction, with the low-surface-brightness emission
extending out 8\,arcsec, or $\sim 24$\,kpc after correcting for
lensing.  This is evidence that the radio emission may be fueled by
supernova activity related to an extended starburst region, rather
than by an AGN, although the relative contributions of L1 and L2
cannot be determined reliably. The extent of the radio source
corresponds approximately to that of the Ly$\alpha$ emission region
(Fig.\,4).

The structure seen within L2 and the AGN-like activity of L1 discussed
below are consistent with L2 being a remnant or tidal debris
associated with an interaction that triggered the activity in L1.

\subsection{Spectral classification}

The optical spectrum of L1 in Fig.\,3 shows a wide range of
high-ionization emission lines that are typical of both AGN and
starburst galaxies. The rest-frame width of the lines is typically
only $\ls 1000$--1500\,km\,s$^{-1}$ (Table\,2); Ly$\alpha$ and N\,{\sc
v} are well separated, as are C\,{\sc iv} and He\,{\sc ii}. The line
profiles are only very slightly asymmetric (Fig.\,3), and the
strongest lines show hints of red wings.  Both of these features
suggest that the narrow line widths are not caused by absorption in
material outflowing from the source. If this was the case, then
significant broad absorption lines (BALs) would be visible on the blue
side of the lines. Weak BAL troughs are visible on the C\,{\sc iv},
Si\,{\sc ii} and Si\,{\sc iv} lines, and outflow velocities of $\sim
5000$\,km\,s$^{-1}$ are indicated.

With the exception of the broad, semi-forbidden C\,{\sc
iii}]\,1909\AA\ line, the narrowness of the emission lines and the
presence of high-ionization lines, such as N\,{\sc v} and C\,{\sc iv},
indicates that the galaxy is best classified either as a very luminous
Seyfert~2 or as one of the rare class of narrow-line or type-2 QSOs
(Baldwin et al.\ 1988), with weak BAL features. Using the definition
of Heckman et al.\ (1995), based on the ratio of the strengths of the
C\,{\sc iv} and He\,{\sc ii} lines, the galaxy would be placed in the
narrow-line QSO class.

However, a detailed classification of L1 is unwarranted, especially
given the presence of both narrow permitted and broad semi-permitted
lines in a single galaxy. For our purposes it is enough to know that
the characteristics of both Seyfert~2 and narrow-line QSOs are believed
to arise from a dusty AGN, with emission from both a slightly obscured
narrow-line region and a highly obscured broad-line region, perhaps
combined with a circum-nuclear star-burst.  In the permitted lines,
emission from the narrow-line region dominates, exhibiting a range of
high-excitation line emission, with FWHM~$\sim 1000$\,km\,s$^{-1}$.
In the case of L1, the narrow-line region is seen through a weak
BAL outflow.  Some light does escape from the broad-line region and
is seen in the semi-forbidden lines, such as C\,{\sc iii}]\,1909\AA\
which exhibits a broad component. Hard X-rays may also escape (Ohta et
al.\ 1996).  The absence of a strong broad component to the H$\alpha$
line in L1 indicates that the reddening to the broad-line region of this
galaxy, $A_V$, probably exceeds $\sim 8$--10.

The spectrum of L1 shares many of the same properties shown in the
well-studied {\em IRAS} source F\,10214+4724 (e.g.\ Serjeant et al.\
1998), which has been variously described as a proto-galaxy, starburst
or Seyfert~2.  Both SMM\,02399$-$0136 and F\,10214+4724 show
predominantly narrow emission lines, with broad C\,{\sc
iii}]\,1909\AA\ and low ratios of Ly$\alpha$ to N\,{\sc v} (Serjeant
et al.\ 1998).

Turning to the spectrum of L2, we see a stronger Ly$\alpha$ compared
to the continuum, than in L1; also, N\,{\sc v}, C\,{\sc iv} and
He\,{\sc ii} are absent. The morphology of L2 is suggestive of an
interaction or merger between L1 and a companion, rather than as
extended QSO `fuzz' (e.g.\ Hutchings 1995). There are, however,
spectral similarities between L2 with the extended emission-line
regions associated with radio-loud quasars (e.g., Lehnert \& Becker
1998); it is interesting to speculate that since the tangential
amplification experienced by L2 leads us to postulate that it is the
remnant of a merger, perhaps many examples of `fuzz' have similar
origins.

\subsection{Spectral energy distribution}

%
%
\begin{figure*}
\centerline{\psfig{file=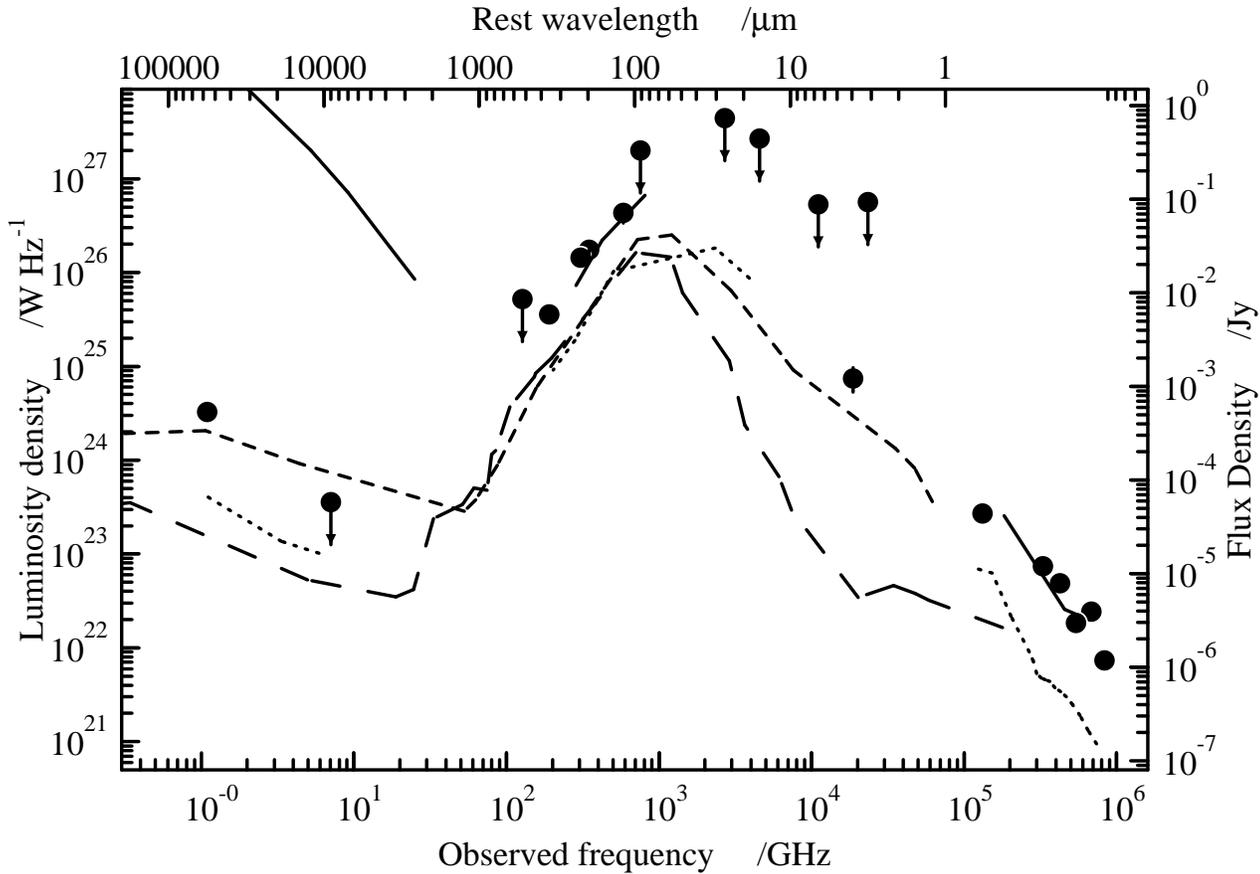,width=15.0cm}}
\caption{The SED of SMM\,02399$-$0136 between the radio and optical
wavebands, represented by filled circles. The right-hand scale gives
the flux densities for this galaxy. For comparison, we have plotted
the SEDs of the ultraluminous {\em IRAS} sources F\,10214+4724 (dots),
Mrk~231 (dashes) and Arp~220 (long dashes) and the radio galaxy
8C\,1435+635 (solid), with units of luminosity density (left-hand
scale). These lines are broken in regions where only upper limits to
the SEDs are available. The SEDs were taken from Rowan-Robinson et
al.\ (1993), Barvainis et al.\ (1995), Ivison et al.\ (1997, 1998) and
D.\,H.\ Hughes (private communication); for F\,10214+4724 and
SMM\,02399$-$0136 they are corrected for lensing by factors of 30 and
2.5 respectively.}
\end{figure*}

The radio--UV SED of L1 deduced from our observations is shown in
Fig.\,5. At wavelengths longer than about 10\,$\mu$m this is
effectively the SED of both L1 and L2. We have chosen not to confuse
Fig.\,5 by plotting the UV--optical data for L2.  The SEDs of other
far-IR luminous galaxies are also shown in Fig.\,5.  Comparing the
far-IR luminosity of SMM\,02399$-$0136 with that of F\,10214+4724
(corrected for a lens amplification of 30) in Fig.\,5, we see that
SMM\,02399$-$0136 is roughly five times more luminous, although such
comparisons are sensitive to the exact amplification suffered by the
emission regions in F\,10214+4724 probed at these wavelengths, which
is highly uncertain (Broadhurst \& Lehar 1995).  The closest match to
SMM\,02399$-$0136 in the UV and optical wavebands is with
8C\,1435+635, a dusty radio galaxy --- and presumably a massive
elliptical --- at $z=4.25$. 8C\,1435+635 also provides the best match
to the SED in the far-IR waveband, but its radio luminosity far
exceeds that of SMM\,02399$-$0136.  We briefly discuss the properties
of the galaxy across the wavelength range shown in Fig.\,5, starting
at the shortest wavelengths.

Both L1 and L2 have a smooth, steep UV--optical--mid-IR continuum and
blue optical and near-IR colours. The exception being the red $(U-B)$
colour caused by the presence of the redshifted Lyman-limit in the
F336W passband.  L2 is slightly bluer than L1, the UV--optical
spectral index, $\alpha$, where $S_{\nu} \propto \nu^{+\alpha}$, is in
the range $-1.5$ to $-1.9$ for L1 --- much steeper than that seen in
most unobscured QSOs, and closer to $\alpha \simeq -1.1$ for L2.  The
distance modulus at $z=2.8$ is 47.64, and so L1 has a $K$-corrected
absolute magnitude $M_B = -24.0$ for a UV--IR spectral index,
$\alpha=-1.65$, while L2 has $M_B = -22.1$ (with $\alpha \simeq
-1.1$).  Across the rest-frame UV--IR spectral index we find a
similarly steep spectrum with $\alpha = -1.65$ for the composite
source L1/L2, although the SED is poorly constrained at wavelengths
between about 25 and 50\,$\mu$m. {\em ISO} observations probing these
wavelengths will be crucial for the accurate determination of the
temperature and optical depth of the dust that clearly dominates the
SED at longer wavelengths (Frayer et al.\ 1998, in preparation).

In the far-IR, at rest-frame wavelengths between 120 and 350\,$\mu$m,
observed by SCUBA, the SED has the characteristic spectral index
$\alpha \simeq +3$ of optically-thin emission from dust grains.
Simple fits suggest that the dust temperature $T_{\rm d} \simeq
40$--50\,{\sc k}, although higher temperatures can be fit if the
opacity of the dust becomes significant at wavelengths of about
100--200\,$\mu$m. The luminosity in the far-IR waveband between
observed wavelengths of 20 and 1000\,$\mu$m, $L_{\rm FIR}$, is about
$10^{13}$\,L$_{\odot}$ if $T_{\rm d} = 40$\,{\sc k}, and several times
larger if either the dust opacity is significant or if $T_{\rm d}$
exceeds 40\,{\sc k}. Adopting $k_{\rm d} = 0.15
[\lambda_0/800\,\mu{\rm m}]^{-1.5}$ for the standard dust emission
parameter in order to facilitate comparisons with other distant
galaxies (c.f.\ Ivison et al.\ 1998), where $\lambda_0$ is the
rest-frame wavelength, we predict dust masses $M_{\rm d}= 7\times
10^8$ and $5\times 10^8$\,M$_{\odot}$, for $T_{\rm d} = 40$ and
50\,{\sc k} respectively.

If the dust in SMM\,02399$-$0136 is heated primarily by hot, young
OB-type stars then the far-IR luminosity corresponds to a formation
rate of stars with masses larger than 10\,M$_\odot$ of about
$2000$\,M$_{\odot}$\,yr$^{-1}$, and to a total SFR of $\sim
6000$\,M$_{\odot}$\,yr$^{-1}$ if the initial mass function (IMF)
extends down to much lower masses (Thronson \& Telesco 1986).  A
similarly high estimate of the SFR is given by the H$\alpha$
luminosity from our $K$-band spectrum.  After correcting for lensing,
but not for possible extinction, the combined H$\alpha$ luminosity of
L1 and L2, $L_{{\rm H}\alpha}$, is $\simeq 5\times 10^{37}$\,W, and so
the total SFR is expected to lie in the range
2000--20000\,M$_{\odot}$\,yr$^{-1}$ (where the range includes the
uncertainty in the relation derived by different authors --- for
example: Barbaro \& Poggianti 1997; Kennicutt 1983; Hill et al.\
1994).  By any standards this would be a spectacular starburst.

At still longer wavelengths, the 8.7- and 1.4-GHz measurements yield a
steep index, $\alpha \ls -1.2$, which is typical for an optically-thin
synchrotron source.  It is steeper than the index measured in
ultraluminous {\em IRAS} galaxies such as Arp~220, Mrk~231 and
F\,10214+4724 -- for F\,10214+4724, $\alpha=-0.9$ (Rowan-Robinson et
al.\ 1993) -- and may well be as steep as the integrated emission from
the radio galaxy, 8C\,1435+635.  The steep spectrum may be the
signature of AGN-related emission: a dense interstellar medium can
frustrate the formation of conventional lobes, causing the magnetic
field to be advected up the jet.  However, several scenarios involving
star formation can equally explain the steep radio spectrum: either
ongoing star formation, where cosmic-ray electrons with the highest
energies are quickly lost to regions where the magnetic field is low;
or a recent cessation of star formation which produces an absence of
newly accelerated electrons.  Furthermore, only 2--10 per cent of the radio
emission from spiral disks and nuclear starbursts is due directly to
supernova remnants (SNRs), the dominant contribution being provided by
cosmic rays (D'Odorico, Goss \& Dopita 1982; Helou et al.\
1985). There is no difficulty, therefore, in reconciling the observed
radio emission with that expected from an intense starburst: first,
the extended radio morphology is consistent with a starburst in L1, or
both L1 and L2, rather than emission from a compact AGN; and secondly,
the total SFR of greater than 1000\,M$_{\odot}$\,yr$^{-1}$, derived
from $L_{\rm FIR}$ and $L_{{\rm H}\alpha}$, is consistent with the
supernova rate of 80--400\,yr$^{-1}$ indicated by the radio emission,
especially if the IMF is top heavy.

Due to the modest resolution of our sub-mm/radio maps we are unable to
determine the relative contributions of L1 and L2 to the emission in
these wavebands; however, the extended 1.4-GHz radio emission is
roughly aligned with the orientation of L1/L2 and so at least part,
and perhaps up to half, of the emission might be expected to originate
in L2.  If this is the case, and if the sub-mm dust emission has a
similar distribution on small scales, then L2 might be the first
non-active, sub-mm-selected high-redshift galaxy, and so the best
candidate for a `primeval' galaxy. Sub-mm observations at higher
angular resolution will be crucial for better understanding the nature
of such galaxies.  The forthcoming generation of very sensitive large
mm/sub-mm interferometers (Brown 1996; Downes 1996) will be especially
effective in advancing these studies.

\section{Discussion}

We have presented detailed follow-up observations of the brightest
of sub-mm-selected galaxy from our deep SCUBA survey of seven lensing
clusters (SIB; Smail et al.\ 1998).  These reveal a radio-quiet,
active galaxy at $z=2.8$ with a total bolometric luminosity of
$\gs10^{13}$\,L$_\odot$.  The H$\alpha$ emission from this galaxy and
its SED in the radio and far-IR wavebands are consistent with an SFR
of several 1000\,M$_\odot$\,yr$^{-1}$. Its optical/UV spectrum is most
consistent with that of a narrow-line, dust-obscured AGN, very similar to
F\,10214+4724. It is possible that either star formation or the active
nucleus, or most likely a combination of the two, is responsible for
the high luminosity of SMM\,02399$-$0136. At present, the data do not
allow us to discriminate between the two mechanisms, as also appears to
be the case with F\,10214+4724 (Serjeant et al.\ 1998).

It is interesting to compare the properties of this galaxy with the
characteristics of the population of optically-selected AGN and to
consider the consequences of this detection for future sub-mm surveys.
In seven fields surveyed to date we can place a limit of
$100^{+330}_{-55}$\,deg$^{-2}$ to the surface density of obscured
AGN/starburst galaxies as intrinsically luminous as SMM\,02399$-$0136
at 90 per cent confidence, if we assume that the mean amplification
factor is about 1.3 in these fields and that the population of
background galaxies is unclustered. This density is roughly 60 per
cent of the surface density of optically-selected QSOs brighter than
$B=22$ (Bershady et al.\ 1998), and is also comparable to the surface
density of faint radio galaxies at flux densities of 0.2\,mJy in the
deepest 1.4-GHz surveys (Oort 1987).  The magnitude and radio flux
density limits given here are similar to the properties of
SMM\,02399$-$0136 after correction for lensing.  Thus our observations
of SMM\,02399$-$0136 are consistent with a population of distant,
obscured AGNs with a surface density comparable to that of
optically-selected quasars.  Classical search techniques in the radio
and optical wavebands are relatively insensitive to radio-quiet,
dust-shrouded galaxies such as SMM\,02399$-$0136, and so conclusions
about the abundance and evolution of active galaxies at high redshifts
that are based upon such approaches may have to be reassessed.
Clearly extensive sub-mm surveys are needed to determine the actual
abundance of such galaxies.

A population of highly obscured AGN might also provide a significant
contribution to the diffuse X-ray background (XRB, Georgantopoulos et
al.\ 1996).  A population of galaxies similar to SMM\,02399$-$0136
with a surface density of 100\,deg$^{-2}$, each with a flux density
close to our lensing-corrected $3\sigma$ upper limit of $4 \times
10^{-15}$\,erg\,s$^{-1}$\,cm$^{-2}$ at 0.1--2.0\,keV and a spectral
index $\alpha \sim -0.5$ (Almaini et al.\ 1995) could contribute up to
around 20 per cent of the total spectral intensity of the XRB at
energies close to 1\,keV (Georgantopoulos et al.\ 1996). Such highly
obscured galaxies have been proposed as the origin of the XRB owing to
their flat spectral shape (Almaini, Fabian \& Barcons 1998).

SIB carried out the first deep survey in the sub-mm waveband, and
detected six sources. Here, we have shown that the luminosity of the
brightest of these may contain a contribution from an active nucleus.
What are the consequences for both the conclusions drawn about the
evolution of galaxies by SIB, and the prospects for future surveys based
on this result?  The preliminary analysis of SIB assumed that the energy
source in all six of their detected galaxies was star formation. Clearly,
at least a part of the luminosity of SMM\,02399$-$0136 is produced
by an active nucleus, and we have no further information about the
other five detections. In general, an AGN will have no effect on the
predicted counts or intensity of background radiation in the sub-mm,
but the rate of production of heavy elements in high-mass stars that is
required to power these sources will be reduced. At present, therefore,
the conclusions of SIB are unaffected, but if the other sources in the
sample are also found to be active then this would change, although
there is no compelling evidence to suggest this is likely.

Larger blank-field surveys in the sub-mm waveband are underway, and so
an expanded sample of galaxies such as SMM\,02399$-$0136 should soon
be compiled (Blain \& Longair 1996; Pearson \& Rowan-Robinson 1996;
Hughes \& Dunlop 1998).  In the longer term ESA's 3.5-m {\it
Far-Infrared and Submillimetre Space Telescope (FIRST)} mission
(Pilbratt 1997) is expected to survey an area of about 0.01\,deg$^2$
at 480\,$\mu$m with 35-arcsec resolution to a 1$\sigma$ sensitivity of
around 0.8\,mJy in a 1-hr integration.  The corrected 450-$\mu$m flux
density of SMM\,02399$-$0136 is 28\,mJy, and so {\it FIRST} could
typically detect about 120 comparable galaxies at 5$\sigma$
significance in each hour of integration.  At 850\,$\mu$m large
ground-based mm/sub-mm-wave interferometer arrays (Brown 1996; Downes
1996) should be capable of surveying of order 0.1\,deg$^2$ to a
1$\sigma$ sensitivity of 2\,mJy in a 1-hr integration (Blain 1996a),
and so could detect about 25 galaxies similar to SMM\,02399$-$0136 at
5$\sigma$ significance in each hour of integration. The prospects for
studying the redshifted mid-IR fine-structure line emission from these
galaxies are also excellent (Loeb 1994; Blain 1996b; Stark
1997). Source confusion is not expected to be a problem in these
observations (Blain, Ivison \& Smail 1998). Even if the abundance of
galaxies such as SMM\,02399$-$0136 is an order of magnitude smaller
than that suggested here, samples of many thousands of dust-shrouded
AGN could still be compiled using {\it FIRST} and ground-based
interferometer arrays, independent of the degree of intrinsic or
line-of-sight extinction.
 
\section{Conclusions}

\begin{enumerate} 
\item We have presented multi-wavelength observations of the first
distant, sub-mm-selected, ultraluminous galaxy, SMM\,02399$-$0136,
spanning the spectral regions from the radio to the X-ray.

\item Spectroscopy in the optical and near-IR wavebands indicate that
the galaxy lies at $z=2.80$ and consists of both a compact,
dust-obscured AGN component, L1, and a companion structure with more
extended emission, L2.  The emission line properties of L1 are very
similar to those exhibited by F\,10214+4724. However, we show that the
intrinsic bolometric luminosity of this galaxy is probably a factor of
five times brighter than that of the {\em IRAS} source, F\,10214+4724.

\item The extended, weak radio emission from the galaxy is consistent
with the sub-mm emission arising from {\em both} the compact and
diffuse components, L1 and L2 respectively. Although we conclude that
the emission from L1 arises either partly or wholly from a
highly-obscured AGN, we ascribe the extended emission and structure
visible in L2 to a very vigorous burst of star formation, probably
triggered by an interaction involving L1 and L2.

\item At present, the detection of a single active galaxy in the
sample of six sub-mm sources discovered by SIB does not compromise the
conclusions of that paper. The predicted source counts and background
radiation intensities are unaltered by finding a non-thermal
contribution to the emission from SMM\,02399$-$0136; the density of
metals at the present epoch, should be reduced by a factor of up to
about 20 per cent, but remains in agreement with current observations.

\item Based on these observations we estimate that the surface density
of highly-obscured active galaxies, similar to SMM\,02399$-$0136, at
$0.5<z<10$ is of order 115\,deg$^{-2}$. If forthcoming SCUBA surveys
confirm this abundance of dust-shrouded AGN, then the proposed {\it FIRST}
space mission and large ground-based interferometer arrays will provide
potentially huge samples of dust-shrouded active galaxies.

\item We are currently undertaking a detailed study of the properties
of the other sources from our sub-mm-selected sample of galaxies
(Blain et al.\ 1998). The results will be presented in Smail et al.\
(1998).
\end{enumerate} 

\subsection*{ACKNOWLEDGEMENTS}

RJI and IRS acknowledge the award of PPARC Advanced Fellowships. JPK
acknowledges support from CNRS and from an ESEP (CNRS/RS) grant. We
acknowledge useful conversations and help from Paul Alexander, Omar
Almaini, Chris Done, K.\,S.\ Dwarakanath, Steve Eales, Katherine Gunn,
Andy Lawrence, Haida Liang, Richard McMahon, Leo Metcalfe, Tim Naylor,
Frazer Owen, Max Pettini, Ian Robson and Tom Shanks.

\end{document}